\documentclass[aps,prd,reprint,nofootinbib,longbibliography]{revtex4-1}

\usepackage{amsmath}
\usepackage{mathtools}
\usepackage{graphicx}
\usepackage[normalem]{ulem}
\usepackage[com pat=1.1.0]{tikz-feynman}
\usepackage{tikz}
\usepackage{aas_macros}
\usetikzlibrary{external}
\usepackage{comment}
\usetikzlibrary{arrows}
\usepackage[colorlinks=true, allcolors=blue]{hyperref}

\newcommand\barparena[1]{\overset{%
   \scriptscriptstyle(-)}{#1}}

\begin{document}

\title{Roles of fast neutrino-flavor conversion on the neutrino-heating mechanism of core-collapse supernova}

\author{Hiroki Nagakura}
\email{hiroki.nagakura@nao.ac.jp}
\affiliation{Division of Science, National Astronomical Observatory of Japan, 2-21-1 Osawa, Mitaka, Tokyo 181-8588, Japan}

\begin{abstract}
One of the greatest uncertainties in any modeling of inner engine of core-collapse supernova (CCSN) is neutrino flavor conversions driven by neutrino self-interactions. We carry out large-scale numerical simulations of multi-energy, multi-angle, three-flavor framework, and general relativistic quantum kinetic neutrino transport in spherical symmetry with an essential set of neutrino-matter interactions under a realistic fluid profile of CCSN. Our result suggests that the neutrino heating in the gain region is reduced by $\sim 40\%$ due to fast neutrino-flavor conversion (FFC). We also find that the total luminosity of neutrinos is enhanced by $\sim 30 \%$, for which the substantial increase of heavy-leptonic neutrinos by FFCs are mainly responsible. This study provides evidence that FFC has a significant impact on the delayed neutrino-heating mechanism.
\end{abstract}
\maketitle

{\em Introduction.}---Most massive stars end their lives as catastrophic explosions known as core-collapse supernovae (CCSNe). It is well established that neutrinos are key players in determining success or failure of explosion (see \cite{2019ARNPS..69..253M,2020LRCA....6....4M,2021Natur.589...29B} for recent reviews). Their primary role on explosions is transport of thermal energy from the vicinity of proto-neutron star (PNS) to the gain region where there is net neutrino heating in post-shock region. Recent CCSN simulations have yielded successful explosions more often than not, in which neutrino heating aided by multi-dimensional (multi-D) fluid instabilities deposits enough energy to revive a stalled shock wave.

There remains a large uncertainty in the explosion mechanism, however. The current CCSN model relies on an assumption that neutrino flavor conversion is suppressed by refractive effects of matter. It has been suggested that another refractive effect by neutrino self-interactions can induce instabilities of flavor conversions (see \cite{2010ARNPS..60..569D,2021ARNPS..71..165T,2022Univ....8...94C,2022arXiv220703561R} for reviews). Detailed inspections of the neutrino data obtained by multi-D CCSN simulations suggested that fast neutrino-flavor conversion (FFC) can ubiquitously occur in CCSN core \cite{2021PhRvD.103f3033A,2021PhRvD.104h3025N}. Since the timescale of the flavor conversion can be shorter by several orders than dynamical timescale, FFC has the power to break the neutrino-matter equilibration in optically thick region \cite{2022arXiv220704058S}. This exhibits a possibility of radical change of both neutrino- and fluid dynamics.

In this {\it Letter}, we present large-scale numerical simulations of multi-energy, multi-angle and general relativistic quantum kinetic neutrino transport with essential neutrino-matter interactions under a fluid background taken from one of our CCSN simulations. We quantify the impact of FFC on neutrino cooling and heating in the optically thick and thin (or gain) regions, respectively. In this study, we pay a special attention to astrophysical aspects in roles of FFC on CCSNe. Detailed discussions of FFC properties will be deferred to another paper. We work in units with $c = G = \hbar = 1$, where $c$, $G$, and $\hbar$ are the light speed, the gravitational constant, and the reduced Planck constant, respectively.

{\em Method and model.}---We solve a spherically symmetric quantum kinetic equation (QKE) for neutrino transport by GRQKNT code \cite{2022PhRvD.106f3011N}. General relativistic effects are taken into account in one of our models by assuming Schwarzschild spacetimes, which is a reasonable approximation outside of a PNS. Assuming spherically symmetry and ultrarelativistic neutrinos, the QKE can be written as (see also Eq.~15 in \cite{2022PhRvD.106f3011N}),
\begin{equation}
  \begin{split}
& \frac{\partial }{\partial t} \biggl[ \Bigl(1 - \frac{2M}{r} \Bigr)^{-1/2} \barparena{f} \biggr]
+ \frac{1}{r^2} \frac{\partial}{\partial r} \biggl[ r^2 \cos \theta_{\nu} \Bigl(1 - \frac{2M}{r} \Bigr)^{1/2}   \barparena{f} \biggr] \\
& - \frac{1}{\nu^2} \frac{\partial}{\partial \nu} \biggl[ \frac{M}{r^2} \Bigl( 1 - \frac{2M}{r} \Bigr)^{-1/2} \nu^3 \cos \theta_{\nu} \barparena{f} \biggr] \\
&- \frac{1}{\sin \theta_{\nu}} \frac{\partial}{\partial \theta_{\nu}} \biggl[ \sin^2 \theta_{\nu} \frac{r-3M}{r^2} \Bigl(1 - \frac{2M}{r} \Bigr)^{-1/2}   \barparena{f} \biggl] \\
& = \barparena{S} - i \xi [\barparena{H},\barparena{f}],
  \end{split}
\label{eq:Sch1DQKE}
\end{equation}
where $t$, $r$, $M$ denote time, radius, and black hole mass, respectively. $\nu$ and $\theta_{\nu}$ represent the neutrino energy and flight angle in momentum space, which are defined in the local orthonormal frame with the timelike unit normal to spatial hypersurfaces. $\barparena{f}$, $\barparena{S}$, and $\barparena{H}$ denote the density matrix of neutrinos, collision term, and Hamiltonian operators associated with neutrino flavor conversion, while the upper bar denotes those for antineutrinos. QKE in flat-spacetimes can be restored by $M=0$ in Eq.~\ref{eq:Sch1DQKE}. In this study, we consider QKE in both two- and three flavor frameworks. The matter potential is ignored, but we leave the vacuum potential with reduced mixing angles. We adopt normal mass hierarchy with squared mass differences of $\Delta m^2 = 2.5 \times 10^{-6} {\rm eV^2}$ for two flavor approximation, and $\Delta m^2_{21} = 7.42 \times 10^{-5} {\rm eV^2} $ and $\Delta m^2_{31} = 2.51 \times 10^{-3} {\rm eV^2}$ for three flavor one, while all mixing angles are assumed to be $10^{-6}$. $\xi$ in the right hand side of Eq.~\ref{eq:Sch1DQKE} denotes an attenuation parameter of Hamiltonian potential. Although attenuating Hamiltonian is unphysical and rather pragmatic, this prescription is necessary to make the simulations tractable. We can discuss realistic features by varying $\xi$ (see also \cite{2022PhRvL.129z1101N,2022arXiv221101398N}).

Neutrino-matter interactions are one of the key ingredients not only for exchanging lepton, energy, and momentum between neutrinos and matter, but also for FFC dynamics itself 
\cite{2021PhRvD.103f3002S,2021PhRvD.103f3001M,2022PTEP.2022g3E01S,2022PhRvD.105d3005S,2022PhRvD.106l3013K}. 
In this study, we consider an essential set of weak processes: electron-capture by free proton and positron capture by free neutron (neutrino emission), their inverse-reactions (neutrino absorption), and also isoenergetic scatterings with nucleons and nuclei (see \cite{2022PhRvD.106f3011N} for the explicit form of each reaction). The reaction rates are computed based on a fluid profile, which is set by referring a result of spherically symmetric CCSN simulation in \cite{2019ApJS..240...38N} at the time snapshot of $300$ ms after bounce for 15 $M_{\odot}$ progenitor computed in \cite{2002RvMP...74.1015W}, where $M_{\odot}$ denotes the solar mass. Thermodynamical quantities and compositions are obtained by a nuclear statistical equilibrium equation-of-state \cite{2017JPhG...44i4001F}.

\begin{figure}
    \includegraphics[width=\linewidth]{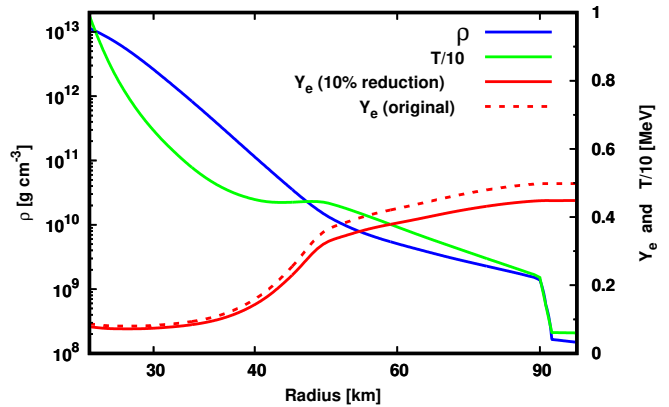}
    \caption{Fluid profile at $300$ ms after bounce for $15 M_{\odot}$ progenitor in a CCSN model of \cite{2019ApJS..240...38N}. We display baryon mass density ($\rho$) and temperature ($T$) as blue and green lines, respectively. We adopt electron-fraction $Y_e$ with $10\%$ reduction (red solid line) from that obtained the original CCSN simulation (red dashed line). See text for more details.
}
    \label{graph_fluidprofile}
\end{figure}

We note that electron-lepton number (ELN) crossing, which is a necessary and sufficient condition to trigger FFCs, is unlikely to occur in the post-shock region in spherically symmetric models. Recent multi-D CCSN simulations, however, commonly suggested that electron-fraction ($Y_e$) at the inner region of neutrino sphere becomes lower than in spherically symmetric models. This is mainly due to fluid instabilities (e.g., PNS convection \cite{2020MNRAS.492.5764N}), asymmetric matter motions accompanied by PNS kick \cite{2019ApJ...880L..28N} or LESA \cite{2014ApJ...792...96T}, and stellar rotation \cite{2022ApJ...924..109H}, which offer a preferable condition for occurrences of ELN crossings \cite{2019ApJ...886..139N,2021PhRvD.104h3025N}. We, hence, adopt $Y_e$ profile as $10 \%$ reduction from that in the spherically symmetric CCSN model, while baryon mass density ($\rho$) and temperature ($T$) profiles are directly taken from the data (see Fig.~\ref{graph_fluidprofile}). We confirm that ELN crossing appears at $\gtrsim 30$km in the simulation of classical neutrino transport (hereafter denoted as no flavor conversion or NFC model). It is also worthy of note that we checked the dependence of $Y_e$ reduction by carrying out simulations with $7 \%$ reduction of $Y_e$. We confirmed that the impact of FFCs on CCSN explosion is similar as the case in the $10 \%$ reduction.

\begin{table}[t]
\caption{Summary of our models.
}
\begin{tabular}{ccccccc} \hline
~~model~~ & spacetime & flavor & ~~$\xi$~~ & ~~$\Delta r_{\rm min}$ [cm]~~ & ~~$N_r$~~ \\
 \hline \hline
M3F & flat & 3 & $10^{-4}$ & 30 & 12,288 \\
M3FGR & BH ($1.5 M_{\odot}$) & 3 & $10^{-4}$ & 30 & 12,288 \\
M2F & flat & 2 & $10^{-4}$ & 30 & 12,288 \\
M3FH & flat & 3 & $2 \times 10^{-4}$ & 15 & 24,576 \\
 \hline
\end{tabular}
\label{tab:model}
\end{table}

\begin{figure*}
    \includegraphics[width=\linewidth]{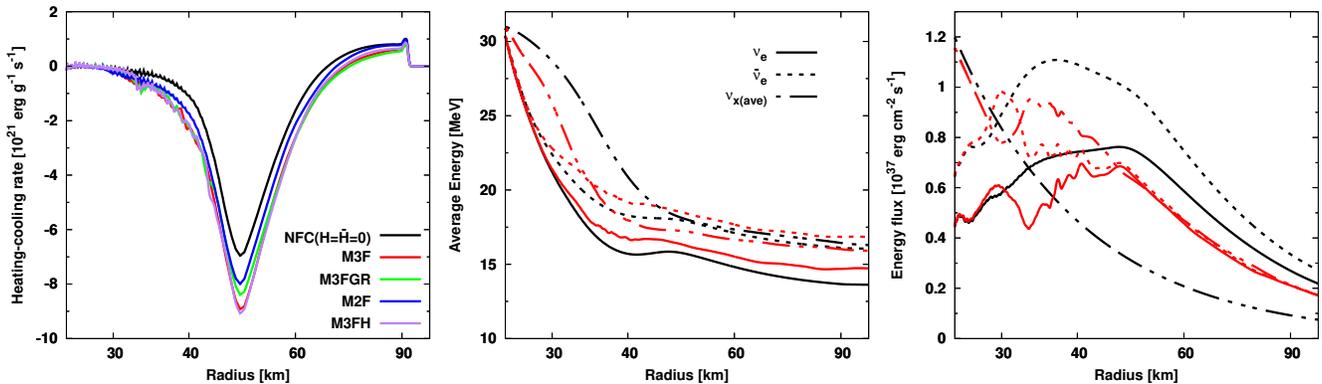}
    \caption{Radial profiles of three key quantities. Left: gain energy from neutrinos. Each color corresponds to a different model. Middle: average energy of neutrinos. Line type distinguishes the species of neutrinos. Right: energy flux of neutrinos.
}
    \label{graph_Heat_Aveene_Eneflux}
\end{figure*}

In this study, we focus on a spatial domain of $20 {\rm km} \le r \le 100 {\rm km}$ which covers from optically thick to thin regions. The radial grid is nonuniform, in which the grid width increases geometrically. We adopt a uniform grid for directional cosine of neutrino flight directions with $96$ grid points. We cover $0 {\rm MeV} \le \nu \le 80 {\rm MeV}$ in the neutrino energy with $12$ grid points. The lowest-energy grid covers from $0 {\rm MeV} \le \nu \le 2 {\rm MeV}$ and the rest of the energy grids is discretized logarithmically.

To set an initial condition, we run two simulations (flat spacetime and black hole one) as NFC models until the system reaches an equilibrium state. At the inner boundary in space, we adopt a Dirichlet boundary condition for outgoing neutrinos by assuming that neutrinos are in equilibrium with matter. For incoming neutrinos, we adopt a free boundary condition. At the outer boundary, we assume no neutrinos in the incoming directions, while we use a free boundary condition for outgoing neutrinos.

In Table~\ref{tab:model}, we summarize numerical setup and some parameters characterizing each model. M3F represents a reference model, in which flat spacetimes ($M=0$) and three flavor framework are assumed. GR effects can be studied from M3FGR with $M=1.5 M_{\odot}$, while other setups are the same as those used in M3F. We also run a simulation with two flavor framework (M2F model). We adopt $\xi = 10^{-4}$ and $N_r=12288$ (the number of radial grid) for all these three models. To check the dependence of attenuation parameter, we run M3FH, in which we set $\xi = 2 \times 10^{-4}$ with higher spatial resolutions; the smallest grid width, $\Delta r_{\rm min}$, is $15$cm and $N_r$ is 24576. We run each simulation for $1$ms and confirm that the system reaches in a quasisteady state. Since we are interested in astrophysical aspects, temporal variations of FFCs are not our focus. We, hence, extend each simulation for $0.05$ms, and all results presented below are computed based on the time-averaged quantities during the time interval.

{\em Results.}---
Left panel in Fig.~\ref{graph_Heat_Aveene_Eneflux} displays radial profiles of net gain energy from neutrinos. As a representative case, we first focus on M3F model. As shown in the panel, the neutrino heating in the gain region becomes remarkably lower than NFC. More quantitatively, the gain radius is increased by $\sim 7 \%$, causing the reduction of baryon mass ($\sim 23 \%$) in the gain region. The local neutrino heating rate is also reduced, resulting in $\sim 48 \%$ reduction of the net gain energy. This suggests that FFC potentially hinders the delayed neutrino-heating mechanism. It may be, however, premature to conclude that FFCs play negative roles on explosions. As shown in the same figure, neutrino cooling in optically thick region is higher in M3F than NFC. Indeed, we find that the total energy flux of neutrinos at the outer boundary is increased by $\sim 33 \%$. This can lead to higher matter temperature due to an efficient contraction of PNS, and therefore the average energy of neutrinos can also be increased, facilitating neutrino absorptions in the gain region. This suggests that feedback from neutrino-matter interactions to fluid dynamics needs to be included to determine whether FFC has a positive or negative role on driving explosion. Its detailed investigation requires radiation-hydrodynamic simulations, which will be addressed in future work.

It is worthy of note that the average energy of electron-type neutrinos ($\nu_e$) and their antipartners ($\bar{\nu}_e$) in M3F become higher than the case with NFC (see middle panel in Fig.~\ref{graph_Heat_Aveene_Eneflux}). This is attributed to the fact that some heavy-leptonic neutrinos ($\nu_x$), having the highest energy among flavors, convert to $\nu_e$ and $\bar{\nu}_e$. On the other hand, energy fluxes of $\nu_e$ and $\bar{\nu}_e$ become lower (see the right panel of Fig.~\ref{graph_Heat_Aveene_Eneflux}), which is also due to lower energy flux of $\nu_x$ in NFC. These two effects compete with each other regarding neutrino heating, and the latter effect dominates over the former. We also find that the energy flux of $\nu_{x(ave)}$, averaging over $\nu_x$ and $\bar{\nu}_x$, are substantially increased in M3F, whereas their average energy becomes lower than in NFC models. This trend is qualitatively in line with results of radiation-hydrodynamic simulations of binary neutron star merger remnant \cite{2022PhRvD.105h3024J,2022PhRvD.106j3003F}.

\begin{figure*}
    \includegraphics[width=\linewidth]{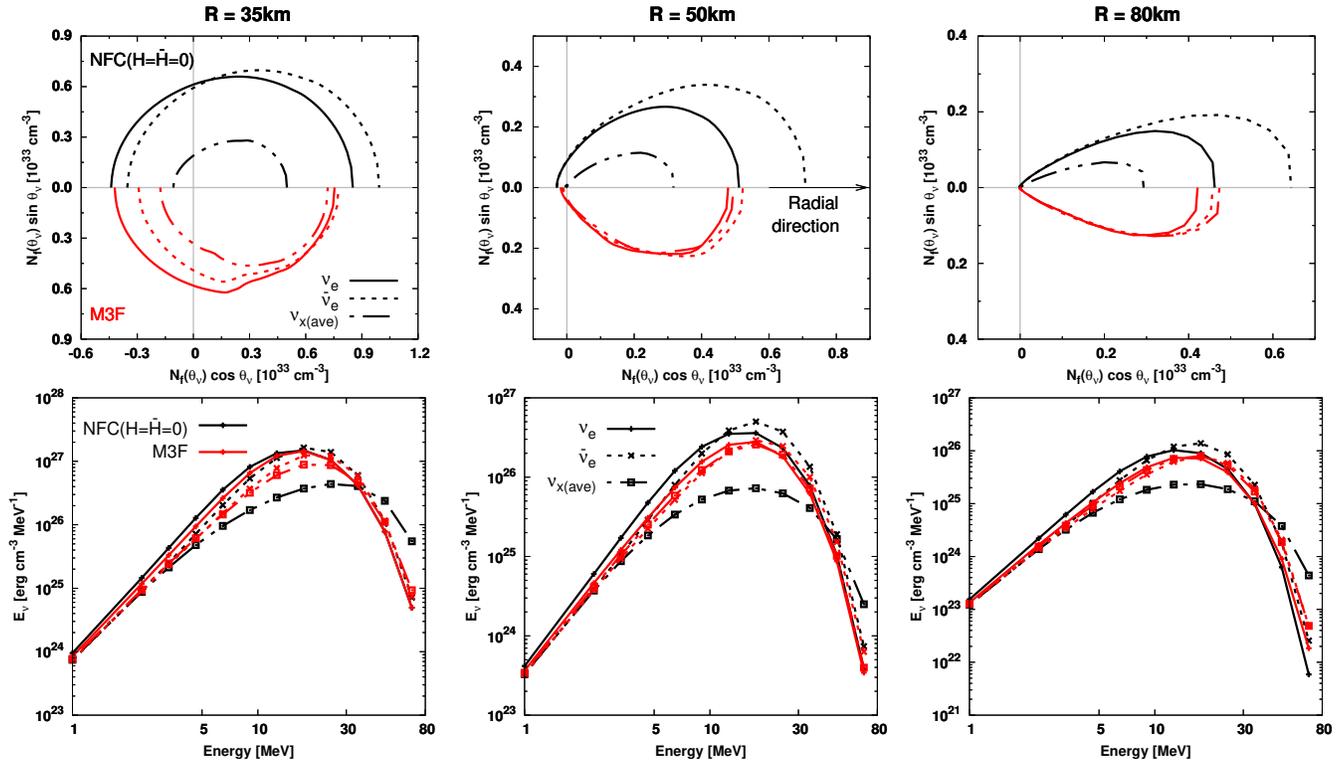}
    \caption{Top: angular distributions of neutrinos at three selected radii: $R=35, 50,$ and $80{\rm km}$ from left to right. Horizontal gray line distinguishes the case with NFC and M3F. The vertical line displays the boundary between outgoing and incoming neutrinos with respect to radial direction (see the arrow in the middle panel). Bottom: Neutrino energy spectra at the three different radii. Color and line style are the same as those used in Fig.~\ref{graph_Heat_Aveene_Eneflux}.
}
    \label{graph_Spect_Ang}
\end{figure*}

We make remarks on model-dependent features on neutrino heating. First, the impact of FFC in M2F is less remarkable than M3F (see in the left panel of Fig.~\ref{graph_Heat_Aveene_Eneflux}); the net gain energy is $\sim 16 \%$ lower than the case with NFC. This indicates that $\nu_e$- and $\bar{\nu}_e$ conversions to heavy-leptonic neutrinos are mild compared to the three flavor framework, which is consistent with the difference of flavor equipartition between these frameworks. Our result exhibits the importance of three flavor framework to quantify the actual impact of FFCs on CCSNe. Next, we find that M3FGR has essentially the same result as M3F, suggesting that GR effects are subdominant. Quantitatively speaking, however, we find neutrino cooling in the semi-transparent region ($\sim 50$km) is suppressed in M3FGR. The lower neutrino cooling exhibits that the number (or energy) density of $\nu_e$ and $\bar{\nu}_e$ is higher than in the NFC model, since the increase of neutrino population leads to larger blocking factor for neutrino emission and also higher neutrino absorption there. The increase of neutrino number is a natural outcome of redshift effect, since the average-energy of neutrinos becomes lower, resulting in the larger neutrino diffusion due to the lower opacity. Finally, we confirm that M3FH model, which has the highest resolution with the modest $\xi$, shows the similar result to M3F, in which the neutrino heating in the gain region is reduced by $\sim 40 \%$ by FFCs.

In Fig.~\ref{graph_Spect_Ang}, we show energy- and azimuthal-integrated angular distributions, $N_{f} (\theta_{\nu}) \equiv \int \int f \varepsilon^2 d \varepsilon d \phi_{\nu} /(2 \pi)^3 $, in the top panel and angular-integrated energy spectra in the bottom panel, for each flavor of neutrinos. Here, we again focus on the result of M3F to discuss key rolls of FFCs in changing neutrino distributions in momentum space. The left panels exhibit that FFC can change both angular distribution and energy spectrum of neutrinos in optically thick region. One thing we do notice here is that an ELN crossing appears at $\cos \theta_{\nu} \sim 0$ in NFC, which guarantees that FFC occurs in M3F. The flavor conversion is vigorous at $\cos \theta_{\nu} \sim 1$, and the flavor equipartition is nearly achieved in the same angular direction. $\bar{\nu}_e$ is reduced more substantially than $\nu_e$, which seems to be due to larger population of $\bar{\nu}_e$ than $\nu_e$ in this direction. For incoming neutrinos ($\cos \theta_{\nu} < 0$), the conversion becomes inefficient, but it is still noticeable for $\nu_{x(ave)}$. The substantial change of $\nu_{x(ave)}$ can also be seen in the energy spectrum, whose feature is strongly dependent on energy. In the high energy region ($\gtrsim 40$MeV), $\nu_{x(ave)}$ in M3F is remarkably lower than NFC, whereas the difference between NFC and M3F is subtle for $\nu_e$ and $\bar{\nu}_e$. This result exhibits that FFC offers a new channel to absorb heavy-leptonic neutrinos, in which $\nu_x$ and $\bar{\nu}_x$ convert to $\nu_e$ and $\bar{\nu}_e$, respectively, by FFCs and then they are absorbed by nucleons via charged-current reactions. In the low energy region ($\lesssim 40$MeV), on the contrary, $\nu_e$ and $\bar{\nu}_e$ emission by charged-current reactions can supply $\nu_x$ and $\bar{\nu}_x$ via FFC. These energy-dependent features of flavor conversion are qualitatively similar as those in collisional instability (see, e.g., \cite{2022arXiv221008254X}).

In semi-transparent ($\sim 50$km) and optically thin ($\sim 80$km) regions, the overall trend of the impact of FFCs on neutrino momentum space is essentially the same as those in the optically thick region (see middle and right panels in Fig.~\ref{graph_Spect_Ang}). One intriguing feature arising in the optically thin region is that $\nu_x$ and their antipartners can be more populated than $\nu_e$ and $\bar{\nu}_e$ (see the right top panel in Fig.~\ref{graph_Spect_Ang}). It is hard to explain this feature only by FFCs (see, e.g., \cite{2021PhRvD.104j3023R,2022PhRvD.106j3039B,2022arXiv221109343Z}), but the interplay among FFC, neutrino-matter interaction, and advection can account for the trend. As we discussed above, $\nu_x$ and $\bar{\nu}_x$ can be more populated in the optically thick region by FFCs. On the other hand, they are more transparent than $\nu_e$ and $\bar{\nu}_e$
due to the lack of charged current reactions. This suggests that the large number of $\nu_x$ and $\bar{\nu}_x$ can diffuse out radially, while the number density of $\nu_e$ and $\bar{\nu}_e$ sharply decreases with radius due to absorptions by matter. As a result, $\nu_x$ can overwhelm $\nu_e$ at the large radii. We also note that there are some quantitative differences between $\nu_x$ and $\bar{\nu}_x$. It is an intriguing question how they have an influence on neutrino signal, although the detailed study is left for our future work.

{\em Summary.}---We present results of general relativistic neutrino transport simulations with taking into account neutrino advection, matter interactions, and flavor conversion under realistic CCSN fluid background. The most striking result is that FFC reduces the efficiency of neutrino heating in the gain region, and that it also accelerates neutrino cooling inside of neutrino sphere by the increase of energy flux of $\nu_x$ and $\bar{\nu}_x$. These effects compete with each other in terms of driving explosion. We postpone the detailed analysis of the competition until radiation-hydrodynamic simulations with QKE neutrino transport are available.

Although qualitative trends found in this study would hold, more systematic studies with varying time snapshots and progenitor models are needed to understand the actual impact of FFCs on CCSN dynamics. \citet{2023arXiv230111938E} recently carried out radiation hydrodynamic simulations of CCSNe under a parametric treatment of FFC, in which intriguing time-dependent features of FFCs emerged. They also showed that FFCs can lead to significant impacts on CCSN explosions, which is in line with our conclusion. As another caveat in the present study, symmetry assumptions, spherically symmetry in real space and axisymmetry in momentum space, employed in this study may affect FFC dynamics. It is also an intriguing question how many-body neutrino correlations change dynamics of FFCs, which is currently one of the active areas of neutrino research (see, e.g., a recent review \cite{2023arXiv230100342P}). These remaining issues notwithstanding, our result provides evidence that FFC is a key ingredient in the delayed neutrino-heating mechanics, albeit possibly negative impact.

\section{Acknowledgments}
H.N is grateful to Masamichi Zaizen, Chinami Kato, Lucas Johns, and Shoichi Yamada for useful comments and discussions. This work is supported by the HPCI System Research Project (Project ID: 220173, 220047, 220223, 230033), XC50 of CfCA at the National Astronomical Observatory of Japan (NAOJ), Yukawa-21 at Yukawa Institute for Theoretical Physics of Kyoto University, and the High Energy Accelerator Research Organization (KEK). For providing high performance computing resources, Computing Research Center, KEK, and JLDG on SINET of NII are acknowledged. H.N is also supported by Grant-inAid for Scientific Research (23K03468).
\bibliography{bibfile}

\end{document}